\documentclass[a4paper]{article}
\usepackage{INTERSPEECH2021}
\usepackage{multirow}
\usepackage{booktabs}
\usepackage{caption}
 
\usepackage[colorlinks,linkcolor=blue]{hyperref} 
 
\title{Beijing ZKJ-NPU Speaker Verification System for VoxCeleb Speaker Recognition Challenge 2021}
\name{Li Zhang$^1$, Huan Zhao$^1$, Qinglin Meng$^2$ , Yanli Chen$^2$, Min Liu$^2$, Lei Xie$^1$ }
 
\address{
  $^1$Audio, Speech and Language Processing Group (ASLP@NPU), School of Computer Science, Northwestern Polytechnical University (NPU), Xi'an, China \\
  $^2$Beijing ZKJ Technology Co., Ltd. }
\email{lizhang.aslp.npu@gmail.com, zhaohuan@mail.nwpu.edu.cn, lxie@nwpu.edu.cn}

\begin{document}
\maketitle 
\begin{abstract}
In this report, we describe the Beijing ZKJ-NPU team submission to the VoxCeleb Speaker Recognition Challenge 2021 (VoxSRC-21). We participated in the fully supervised speaker verification track~1 and track~2. In the challenge, we explored various kinds of advanced neural network structures with different pooling layers and objective loss functions. In addition, we introduced the ResNet-DTCF, CoAtNet and PyConv networks to advance the performance of CNN-based speaker embedding model. Moreover, we applied embedding normalization and score normalization at the evaluation stage. By fusing 11 and 14 systems, our final best performances (minDCF/EER) on the evaluation trails are 0.1205/2.8160\% and 0.1175/2.8400\% respectively for track~1 and 2. With our submission, we came to the second place in the challenge for both tracks.
\end{abstract}


\section{Introduction}
As in previous years~\cite{chung2019voxsrc,nagrani2020voxsrc}, VoxSRC-21 has four challenge tracks, which are fully supervised speaker verification with VoxCeleb~2 development dataset~(track~1), fully supervised speaker verification with arbitrary training set~(track~2), self-supervised speaker verification with VoxCeleb~2 development set~(track~3) and speaker diarisation~(track~4). Our team has worked on track~1 and track~2 -- the fully supervised task particularly focusing on multi-lingual data. Totally we trained 15 systems based on multiple variants of TDNN ~\cite{snyder2018x} and ResNet~\cite{he2016deep}. Moreover, we explored large margin fine-tune strategy, embedding normalization and score normalization to further improve the results.   

\section{Data}

\subsection{Training and Evaluation Datasets}
In this paper, the VoxCeleb~2 development set is the only data set used to train all the models for both tracks, though other data sets are allowed for track 2. We evaluated all our models on the VoxSRC-21 development trials~(VoxSRC21-dev) while the model fusion submissions are evaluated on the VoxSRC-21 test trials~(VoxSRC21-test) due to the limitation on the number of uploading scores.

\subsection{Data Augmentation}
Online data augmentation~\cite{cai2020fly} is performed for all speaker embedding models. Specifically, frequency-domain specAug, additive noise augmentation and reverberation augmentation are adopted for all models, and some models further go through the speaker augmentation step.

\begin{itemize}
    \item \textbf{Frequency-Domain SpecAug:}
We apply time and frequency masking as well as time warping to the input spectrum (frequency-domain 
implementation)~\cite{park19e_interspeech}
    \item \textbf{Additive Noise:} We add the noise, music and babble type from MUSAN~\cite{snyder2015musan} to the original speech of VoxCeleb~2. 
    \item \textbf{Reverberation:} Reverberation is convoluted with the original speech in VoxCeleb~2~\cite{snyder2018x} and the RIRs are from~\cite{habets2006room}.
    \item \textbf{Speaker-Aug:}
Speaker-Aug is a successful method in speech and speaker recognition tasks~\cite{yamamoto2019speaker}. We use the SOX toolkit to perturb speech speed into 0.9 or 1.1 times. The generated utterances are considered as new speakers. As a result, the whole speaker number after speaker-aug becomes 5,994 $\times$ 3 = 17982.
\end{itemize} 


\section{Feature Extraction}
In this challenge, we train the speaker embedding models with Mel Frequency Cepstral Coefficents~(MFCCs) and Mel-Filter bank~(Fbank) features. Specifically, E-TDNN used 60-dimensional MFCCs as the input features. Others used 80- or 96-dimension Fbank as the input features instead.

\section{Deep Speaker Embedding Models}
In total, we trained 15 advanced speaker embedding models and all are variants of x-vcector~\cite{snyder2018x} and ResNet~\cite{he2016deep}. We introduce them in details in the following.
\subsection{E-TDNN}
The extended TDNN architecture~(E-TDNN) \cite{snyder2019speaker} has slightly wider temporal context and interleaves dense layers between convolutional layers compared with the original x-vector model~\cite{snyder2018x}. We adopt the same structure as~\cite{snyder2019speaker}. It comprises 4 blocks of 1D dilated convolutions plus affine layers. The first layer uses a kernel of size 5 and the other ones use a kernel of size 3. The dilation factors are 1, 2, 3 and 4 respectively. The embedding is extracted from the output of an affine layer that follows the statistics
(mean + standard deviation) pooling layer~\cite{garcia2020jhu}. This model is trained by the cross entropy loss.  

 
\subsection{ECAPA-TDNN}
ECAPA-TDNN is known as one of the state-of-the-art speaker verification models~\cite{desplanques20_interspeech}. We train particularly two ECAPA-TDNN models, which configured with 1024 channels and 2048 channels respectively~\cite{thienpondt2021idlab}. The pooling layer of ECAPA-TDNN~(1024) is attention statistic pooling~\cite{desplanques20_interspeech}. 

\subsection{ResNet}
We train 7 ResNet34-related models with the similarity structure but different attention modoul. They have \{64, 128, 256, 512\} or \{32, 64, 128, 256\} channels of residual blocks and multi-head attention statistic pooling. The loss functions are additive angular margin softmax (AAM-softmax), subcenter additive angular margin softmax (SC-AAM-softmax) and circle loss respectively.  
\subsection{ResNet-SE}
ResNet with squeeze and excitation attention (ResNet-SE) has achieved good performance in speaker verification~\cite{thienpondt2021idlab,zhang21g_interspeech} recently. In this work, we adopt ResNet34-SE with 256 channels of SE attention modules. Two ResNet-SE models with different size are particularly adopted. Their channel configurations of residual block are \{32, 64, 128, 256\} and \{64, 128, 256, 512\} respectively~\cite{heo2020clova}. The loss functions of ResNet34-SE~(256) and ResNet34-SE~(512) are AAM-softmax and SC-AAM-softmax respectively.

\subsection{ResNet-BAM}
Our recent study has shown that adding bottleneck attention modules~(BAM) in ResNet leads to improved speaker verification performance~\cite{zhang20ca_interspeech}.  BAM is able to emphasize important elements in 3D feature map generated from convolution. Specifically for the speech signal, 3D feature map has channel dimension (filter number of convolution), time dimension as well as frequency dimension. We adopt the ResNet-BAM model in our work and the channels of each residual block are \{64, 128, 256, 512\}.  

\subsection{SE-Res2Net}
In order to capture speaker characteristics at different levels, we use the Res2Net \cite{2019Res2Net} network with the SE module. The channel configurations of each residual block are \{32, 64, 128, 256\}. In addition, we replace the last two fully connected (FC) layers of Res2Net with a TDNN layer to better model the context information.
\subsection{D-TDNN-SE}
D-TDNN-SE is also adopted, in which dense connectivity~\cite{yu2020densely} is adopted and feed forward and TDNN layers are used to replace the two-dimensional convolutional layers.
\subsection{ResNet-TDNN}
We replace the last linear layer of ResNet with a 4-layer TDNN in order to capture the relationships of different frames, resulting in the ResNet-TDNN model. The number of channels in ResNet is still \{64, 128, 256, 512\}.
\subsection{CNN-ECAPA}
The hybrid CNN-ECAPA~\cite{thienpondt21_interspeech} structure is recently proposed with the belief that the advantages of the convolutional neural network and ECAPA-TDNN are integrated in a unified structure. Specifically, in our CNN-ECAPA, the number of channels in the convolutional stem is set to 128. The output feature map of the convolutional stem is subsequently flattened in the channel and frequency dimensions and used as input for the regular ECAPA-TDNN network.

\subsection{ResNet-DTCF}
We propose the Duality-Temporal-Channel-Frequency (DTCF) attention in ResNet34, named as ResNet-DTCF, to boost the representation extracting capability of CNN in speaker verification. Different
from other squeeze and excitation~(SE) attention learning after averaging the
time and frequency dimensions simultaneously, the DTCF attention module firstly re-calibrates the channel-wise features
with aggregation global context information on temporal and
frequency dimensions, and then the duality channel-wise attention is adopted with preserving temporal and frequency information respectively. The DTCF attention module particularly encodes the temporal and frequency information into the channel-wise attention masks, averting the leakage of global context information in temporal and frequency dimensions. Details of this model can be found from our recent submission \cite{zhangdct} to ASRU2021.
The channels of residual blocks in ResNet-DTCF are \{32, 64, 128, 256\}. 
\subsection{ResNet-Pyramid}
Pyramidal convolution (PyConv) contains a pyramid of kernels, where
each level involves different types of filters with varying size and depth. This design aims to capture different levels of details in speech utterances~\cite{duta2020pyramidal}. Thus in this work, we embed the PyConv into ResNet50 to make the model to capture long contextual information.  The configuration of ResNet-Pyramid is the same as PyConvResNet-50 in~\cite{duta2020pyramidal}. 

\subsection{CoAtNet}
Transformer is a popular sequence-to-sequence model, originally used for neural machine translation (NMT)~\cite{vaswani2017attention} and later introduced to speech recognition~\cite{gulati20_interspeech}. The use of transformer family has become popular in computer vision (CV) recently. We borrow the new CoAtNet~\cite{dai2021coatnet} structure from CV and use it in this work for speaker verification. The oracle transformers, heavily adopting self-attention layers to model global dependencies, have larger model capacity, but their generalization ability can be worse than convolution networks due to the lack of the right inductive bias. The design of CoAtNet~\cite{dai2021coatnet} specifically migrates the problem by combining the advantages of both structures. We explore the performance of CoAtNet in this challenge. The pooling layer of CoAtNet in our model is average pooling and the loss function is cross entropy.

\section{Pooling Layer}
The pooling layer in speaker verification aims to aggregate the frame-level speaker embedding into an utterance-level speaker embedding for scoring. In this challenge, we adopt 5 statistic pooling strategies: temporal average pooling~(TAP)~\cite{yu2014mixed}, statistic pooling~(SP) ~\cite{snyder2018x}, self-attention pooling~(SAP)~\cite{cai18_odyssey}, attentive statistic pooling~(ASP)~\cite{okabe18_interspeech} and multi-head attention pooling~(MHAP)~\cite{india19_interspeech}. The specific use of the pooling layers in different models is summarized in Table~\ref{tab:results}.

\section{Objective Loss Function}
Softmax is the commonly used loss function in classification tasks.
As compared with the softmax loss, the additive angular margin loss (AAM-Softmax)~\cite{liu19f_interspeech} is more popular in speaker verification as increasing intra-speaker distances and ensuring inter-speaker compactness are both important. In this work, we adopt softmax and AAM-Softmax in different models, where $s=30$ and $m=0.2$ are used for AAM-Softmax. Moreover, we also train some models with suncenter AAM softmax~(SC-AAM-softmax) \cite{deng2020sub} and circle loss~\cite{sun2020circle} as well. We particularly introduce the two losses in the following.  Again, the details of the use of different losses can be found in Table~\ref{tab:results}. 

\subsection{SC-AAM-Softmax}
Subcenter additive angular margin loss (SC-AAM-Softmax)~\cite{deng2020sub} adds $K$ sub-centers for each class to reduce the possible impact from label noise in the training data and improve the model robustness. In this challenge, we first set $K$ to 3 and then change it to 1 after the model achieves good result. In this case, the SC-AAM-Softmax has become an ordinary AAM-Softmax.

\subsection{Circle Loss}
A majority of loss functions simply have set an equal penalty strength on within-class similarity ($s_p$) and between-class similarity ($s_n$). On the contrast, circle loss~\cite{sun2020circle} flexibly uses weight factors $\alpha_p$ and $\alpha_n$ to make $s_p$ and $s_n$ learn at different paces. In this work, we set margin to 0.25 and gamma to 64.
 
\section{Training Strategy}
The model training process is composed of base training and fine-tuning. All model are first trained using the Adam optimizer \cite{kingma2014adam} with a cyclical learning rate (CLR) using the triangular2 policy as described in~\cite{smith2017cyclical}. The max and min learning rates are set at $1e-3$ and $1e-8$ respectively.
Then we use the large margin fine-tune (LMF) strategy~\cite{thienpondt2021idlab} to fine-tune our well-trained models. Here the max cyclical learning rate is reduced to $1e-4$. The initial margin penalty and margin of the AAM-softmax layer is set
to 0.2 and 30 respectively. In the fine-tune stage, the penalty and margin are set to 0.25 and 35 instead. In primary training, we use 3s utterances to train the models. In fine-tune step, we adopt 6s utterances to train the models.
 
\section{Normalization and Fusion}

\subsection{Embedding Normalization}
We introduce \textit{embedding averaging} and \textit{matrix scoring} as embedding normalization tricks.
The former is to simply average all segments of each utterance to obtain the speaker representation of the utterance, while the later is to score each trial with one score matrix generated from the segments.

\noindent\textbf{Embedding Average (EA)}
We train our models with 3-second utterances in each batch-size. In the test stage, we also split the test utterances into 3-second clips and extract speaker embedding for each segment. Then we average the embeddings of all segments in the utterance and the averaged one is considered as the final speaker embedding of this utterance. Note that the larger margin fine-tuned models with 6-second utterances are tested on 6-second segments of enrollment and test utterances.  

\noindent\textbf{Matrix Score Average (MSA)}
We split each utterance into $n$ segments and then extract speaker embedding for each segment. Thus for one trail, we can formulate a score matrix with size of $\mathbb{R} ^{(n \times n)}$. Then we average all scores to obtain the final score of this trial.

\subsection{Score Normalization}
Most of our models are scored by cosine similarity but we adopt the PLDA scoring~\cite{snyder2018x} in Kaldi~\cite{povey2011kaldi} when our models trained by the softmax loss function. According to our past experience, PLDA will not work well if the model is trained with margin-based softmax. In addition, we adopt adaptive symmetrical score normalization~(as-norm) to normalize all scores.

\subsection{System Fusion}
System fusion aims to further boost the performance by integrating multiple models which are expected to be complimentary. In the system fusion stage, we adopt manual calibration as well as automatic calibration. According to the performance on the development data set, we adopt the score level fusion that assigns different weights to different models. Considering that the model may over-fit on the development set with manual calibration, we particularly use the BOSARIS toolkit~\cite{brummer2013bosaris} for score calibrating before system fusion.

\section{Experimental Results} 
We evaluate the 15 models with the above mentioned strategies (each with a system index) on voxSRC21-dev and voxSRC21-eval data sets and the results are summarized in Table~\ref{tab:results}. Here ECAPA-TDNN is specifically used for ablation study. We can see from Table~\ref{tab:results} that the EER and minDCF (p=0.05) of ECAPA-TDNN are 3.565\% and 0.1833 on the VoxSRC21-dev trails. Further with the large margin fine-tune~(LMF) strategy, we observe 0.634\%/0.0201 reduction on EER/minDCF. At the evaluation stage, the use of embedding average~(EA) achieves 0.299\%/0.0181 EER/minDCF reduction. In addition, the combination of  matrix score average (MSA) and EA further obtains 0.061\%/0.0215 reduction on EER/minDCF. These results show the effectiveness of various tricks. We also notice that the best single model is ResNet-TDNN~(512) with SC-AAM-Softmax and speaker augmentation (indexed with L2), which achieves the lowest EER/minDCF of 1.614\%/0.0930 on VoxSRC21-dev trials among all our models. As we introduced before, we perform frequency specAug, additive noise and reverberation augmentation dynamically during model training. The speaker augmentation only operates on ECAPA-TDNN (2048), ResNet34-SE (512), CNN-ECAPA, SE-Res2Net, ResNet34-TDNN and CoAtNet. Comparing L1 with M1 -- the only difference of the two systems is the loss function adopted, we can see that SC-AAM-Softmax is superior to circle loss.   

We tried two fusion strategies. In Fusion1 in Table ~\ref{tab:results}, model weights are assigned by hands according to the performance on development trials. In Fusion2-4, models are fused by the BOSARIS toolkit~\cite{brummer2013bosaris} in which a linear fusion strategy is used. In general, system fusion leads to clear performance gain due to the complimentary between different models. Finally on the voxSRC21-eval set, the best performances on track~1 and track~2 are achieved by the fusion of 11 and 14 systems respectively. Due to time limitation, we did not manage to submit a 14-system fusion for track~1. In details, our best minDCF is 0.1205 and 0.1175 for track 1 and 2 respectively, achieving the second place in both tracks.

\begin{table*}[hth]
\caption{The experimental results on VoxSRC21-dev \& VoxSRC21-eval (LMF: large margin fine-tune, EA: embedding average, MSA: matrix score average, SP: statistics pooling, ASP: attentive stitistics pooling, SAP: self-attention pooling)}
\label{tab:results}

\resizebox{\linewidth}{!}{
\begin{tabular}{clccccccc}
\hline
\multirow{2}{*}{\textbf{System Index}} & \multicolumn{1}{c}{\multirow{2}{*}{\textbf{Model Structure}}} & \multirow{2}{*}{\textbf{Pooling Layer}} & \multirow{2}{*}{\textbf{Loss Function}} & \multirow{2}{*}{\textbf{Speaker Aug}} & \multicolumn{2}{c}{\textbf{VoxSRC21-dev}} & \multicolumn{2}{c}{\textbf{VoxSRC21-eval}}  \\ \cline{6-9} 
                                       & \multicolumn{1}{c}{}                                          &                                         &                                         &                                       & \textbf{EER(\%)}     & \textbf{minDCF}\textbf{$_{0.05}$}    & \textbf{EER(\%)}     & \textbf{minDCF$_{0.05}$}      \\ \hline
\textbf{A1}                            & E-TDNN                                                        & SP                                      & Softmax                                 & $\times$                              & 7.384                & 0.4486             &        -            &     -                 \\
\textbf{A2}                            & ~~+PLDA                                                         &  SP                                      &     Softmax                                    & $\times$                                     & 5.562                & 0.3029             &     -                 &   -                   \\
\textbf{B1}                            & ECAPA-TDNN(1024)                                              &       ASP                                     & AAM-Softmax                             & $\times$                              & 3.565                & 0.1833             &        -             &    -                  \\
\textbf{B2}                            & ~~+LMF                                                          & ASP                    & AAM-Softmax                    & $\times$                  & 2.931                & 0.1632             & - & - \\
\textbf{B3}                            & ~~+LMF+EA                                                           & ASP                    & AAM-Softmax                    & $\times$                  & 2.632                & 0.1451             & - & - \\
\textbf{B4}                            & ~~+LMF+EA+MSA                                                   &   ASP                                     &     AAM-Softmax                                   &    $\times$                                  & 2.571                & 0.1236             &      -              &   -                  \\
\textbf{C1}                            & ECAPA-TDNN(2048)                                              & ASP                                     & SC-AAM-Softmax                             & $\checkmark$                              & 3.126               & 0.1511             & - & - \\
\textbf{C2}                            & ~~+LMF+EA+MSA                                                   & ASP                    & SC-AAM-Softmax                    & $\checkmark$                 & 2.012                & 0.1028             & - & - \\
\textbf{D1}                            & ResNet34-SE(256)                                              & SAP                                     & AAM-Softmax                             & $\times$                              & 3.419                & 0.2061             &  -                   &   -                  \\
\textbf{D2}                            & ~~+LMF+EA+MSA                                                   &    ASP                                    & AAM-Softmax                                       &  $\times$                                    & 2.976                & 0.1472             &    -                 & -                   \\
\textbf{E1}                            & ResNet34-SE(512)                                              & SAP                                     & SC-AAM-Softmax                          & $\checkmark$                              & 3.359                & 0.1903             & -                   & -                    \\
\textbf{E2}                            & ~~+LMF+EA+MSA                                                   &   SAP                                     &     SC-AAM-Softmax                                   &     $\checkmark$                                 & 2.252                & 0.1105             &  -                   &  -                   \\
\textbf{F1}                            & ResNet34-DTCF                                                 & ASP                                     & AAM-Softmax                             & $\times$                              & 3.565                & 0.2096             &       -              &  -                   \\
\textbf{F2}                            & ~~+LMF+EA+MSA                                                   &     ASP                                   &    AAM-Softmax                                    &  $\times$                                    & 2.573                & 0.1281             &     -                &   -                 \\
\textbf{G1}                            & ResNet34-BAM                                                  & SAP                                     & AAM-Softmax                             & $\times$                              & 3.384                & 0.1931             &       -              &  -                   \\
\textbf{G2}                            & ~~+LMF+EA+MSA                                                   &    SAP                                    &    AAM-Softmax                                   &    $\times$                                  & 2.623                & 0.1331             &  -                   &    -                 \\
\textbf{H1}                            & ResNet-Pyramid                                                & TAP                                     &AAM-Softmax                                 & $\times$                              & 3.562                & 0.2029             &    -                 &      -               \\
\textbf{H2}                            & ~~+LMF+EA+MSA                                                         &         TAP                            &    AAM-Softmax                                 &     $\times$                               & 2.132                & 0.1051             &   -                  &   -                  \\
 
\textbf{I1}                            & CNN-ECAPA                                                     & SAP                                     & SC-AAM-Softmax                          & $\checkmark$                          & 2.837                & 0.1964             &   -                  &   -                  \\

\textbf{I2}                            & ~~+LMF                                                   & SAP                                       &          SC-AAM-Softmax                             &  $\checkmark$                                    & 2.159                & 0.1310             &       -              &  -                   \\
\textbf{I3}                            & ~~+LMF+EA+MSA                                                   &        SAP                                &            SC-AAM-Softmax                            &    $\checkmark$                                 & 1.773                & 0.0979             &   -                  &     -                \\

\textbf{J1}                            & SE-Res2Net                                                    & MHAP                                    & Circle-Loss                             & $\times$                              & 3.427                & 0.1972             &   -                  &   -                  \\
\textbf{J2}                            & ~~+LMF+EA+MSA                                                   &      MHAP                                  &         Circle-Loss                              &    $\times$                                  & 1.912                & 0.1021             &      -               &     -               \\
\textbf{K1}                            & SE-Res2Net                                                    & MHAP                                    & SC-AAM-Softmax                          & $\checkmark$                              & 3.345                & 0.1824             &       -              &     -                \\
\textbf{K2}                            & ~~+LMF+EA+MSA                                                   &             MHAP                           &             SC-AAM-Softmax                           &   $\checkmark$                                   & 2.017                & 0.1185             &     -               &     -                \\
 
\textbf{L1}                            & ResNet34-TDNN                                                 & MHAP                                    & SC-AAM-Softmax                          & $\checkmark$                              & 3.012                & 0.1532             &      -               &     -                \\
\textbf{L2}                            & ~~+LMF+EA+MSA                                                   &      MHAP                                  &              SC-AAM-Softmax                          &     $\checkmark$                                 & \textbf{1.614}                & \textbf{0.0930}             &       -              &      -               \\
\textbf{M1}                            & ResNet34-TDNN                                                 & MHAP                                    & Circle-Loss                             & $\checkmark$                              & 2.817                & 0.1975             &      -               &     -               \\
\textbf{M2}                            & ~~+LMF+EA+MSA                                                   &        MHAP                                &        Circle-Loss                               &  $\checkmark$                                    & 1.799                & 0.0854             &     -                &   -                  \\
\textbf{N1}                            & D-TDNN-SE                                                     & SAP                                     & AAM-Softmax                             & $\times$                              & 3.424                & 0.2201             &     -                &   -                 \\
\textbf{N2}                            & ~~+LMF+EA+MSA                                                   &        SAP                                &              AAM-Softmax                          &     $\times$                                 & 2.042                & 0.1087             &    -                 &    -                 \\
\textbf{O1}                            & CoAtNet                                                       & SAP                                     & Softmax                                 & $\checkmark$                          & 4.562                & 0.2531             &   -                 &    -                 \\
\textbf{O2}                            & ~~+PLDA                                                   &          SAP                              &               Softmax                         &    $\checkmark$                                  & 2.125                & 0.1814             &    -                 &    -                 \\ \hline
\textbf{Fusion1}                       & \multicolumn{4}{c}{{[}B4+D2+F2+G2+E2{]}}                                                                                                                                               & 1.6384               & 0.0935             & 3.3360               & 0.1659               \\
\textbf{Fusion2}                       & \multicolumn{4}{c}{{[}C2+E2+H2+I2+K2+L2+N2{]}}                                                                                                                                & 1.4384               & 0.0735             & 2.9440               & 0.1457               \\
\textbf{Fusion3}                       & \multicolumn{4}{c}{{[}B4+C2+D2+E2+F2+G2+J2+K2+L2+M2+N2{]}}                                                                                                                                & \textbf{1.4210}                & \textbf{0.0681}             & \textbf{2.8160}               & \textbf{0.1205}               \\
\textbf{Fusion4}                       & \multicolumn{4}{c}{{[}B4+C2+D2+E2+F2+G2+H2+I3+J2+K2+L2+M2+N2+O2{]}}                                        
& \textbf{1.1039}               & \textbf{0.0582}             & \textbf{2.8400}               &\textbf{0.1175}              \\ \hline
\end{tabular}}

\end{table*}
\section{Conclusion}
In our submission to VoxSRC-21, we explored various TDNN-based and ResNet-based speaker embedding models on track~1 and 2. We particularly introduced ResNet-DTCF, CoAtNet and ResNet-Pyramid structures. We also used the large margin fine-tune strategy during model training to further improve the performance. Moreover, we specifically took embedding average and embedding matrix score average as normalization tricks at the evaluation stage. The take-home messages are as follows.

\begin{itemize}
    \item Hybrid structures are always superior as compared with individual ResNet and TDNN neural structures.


   \item Embedding normalization and score normalization during evaluation are useful tricks, which lead to clear performance gain.

   \item Fusion on multiple models is beneficial as models may compliment each other.
\end{itemize}

\section{Acknowledgement}
 This work was supported by Beijing ZKJ Technology Co., Ltd.
\bibliographystyle{IEEEtran}

\bibliography{template}

\begin{thebibliography}{10}
\providecommand{\url}[1]{#1}
\csname url@samestyle\endcsname
\providecommand{\newblock}{\relax}
\providecommand{\bibinfo}[2]{#2}
\providecommand{\BIBentrySTDinterwordspacing}{\spaceskip=0pt\relax}
\providecommand{\BIBentryALTinterwordstretchfactor}{4}
\providecommand{\BIBentryALTinterwordspacing}{\spaceskip=\fontdimen2\font plus
\BIBentryALTinterwordstretchfactor\fontdimen3\font minus
  \fontdimen4\font\relax}
\providecommand{\BIBforeignlanguage}[2]{{%
\expandafter\ifx\csname l@#1\endcsname\relax
\typeout{** WARNING: IEEEtran.bst: No hyphenation pattern has been}%
\typeout{** loaded for the language `#1'. Using the pattern for}%
\typeout{** the default language instead.}%
\else
\language=\csname l@#1\endcsname
\fi
#2}}
\providecommand{\BIBdecl}{\relax}
\BIBdecl

\bibitem{chung2019voxsrc}
J.~S. Chung, A.~Nagrani, E.~Coto, W.~Xie, M.~McLaren, D.~A. Reynolds, and
  A.~Zisserman, ``Voxsrc 2019: The first voxceleb speaker recognition
  challenge,'' \emph{1912.02522}, 2019.

\bibitem{nagrani2020voxsrc}
A.~Nagrani, J.~S. Chung, J.~Huh, A.~Brown, E.~Coto, W.~Xie, M.~McLaren, D.~A.
  Reynolds, and A.~Zisserman, ``Voxsrc 2020: The second voxceleb speaker
  recognition challenge,'' \emph{2012.06867}, 2020.

\bibitem{snyder2018x}
D.~Snyder, D.~Garcia-Romero, G.~Sell, D.~Povey, and S.~Khudanpur, ``X-vectors:
  Robust dnn embeddings for speaker recognition,'' in \emph{2018 IEEE
  International Conference on Acoustics, Speech and Signal Processing
  (ICASSP)}.\hskip 1em plus 0.5em minus 0.4em\relax IEEE, 2018, pp. 5329--5333.

\bibitem{he2016deep}
K.~He, X.~Zhang, S.~Ren, and J.~Sun, ``Deep residual learning for image
  recognition,'' in \emph{Proceedings of the IEEE conference on computer vision
  and pattern recognition}, 2016, pp. 770--778.

\bibitem{cai2020fly}
W.~Cai, J.~Chen, J.~Zhang, and M.~Li, ``On-the-fly data loader and
  utterance-level aggregation for speaker and language recognition,''
  \emph{IEEE/ACM Transactions on Audio, Speech, and Language Processing},
  vol.~28, pp. 1038--1051, 2020.

\bibitem{park19e_interspeech}
D.~S. Park, W.~Chan, Y.~Zhang, C.-C. Chiu, B.~Zoph, E.~D. Cubuk, and Q.~V. Le,
  ``{SpecAugment: A Simple Data Augmentation Method for Automatic Speech
  Recognition},'' in \emph{Proc. Interspeech 2019}, 2019, pp. 2613--2617.

\bibitem{snyder2015musan}
D.~Snyder, G.~Chen, and D.~Povey, ``Musan: A music, speech, and noise corpus,''
  \emph{1510.08484}, 2015.

\bibitem{habets2006room}
E.~A. Habets, ``Room impulse response generator,'' \emph{Technische
  Universiteit Eindhoven, Tech. Rep}, vol.~2, no. 2.4, p.~1, 2006.

\bibitem{yamamoto2019speaker}
H.~Yamamoto, K.~A. Lee, K.~Okabe, and T.~Koshinaka, ``Speaker augmentation and
  bandwidth extension for deep speaker embedding.'' in \emph{INTERSPEECH},
  2019, pp. 406--410.

\bibitem{snyder2019speaker}
D.~Snyder, D.~Garcia-Romero, G.~Sell, A.~McCree, D.~Povey, and S.~Khudanpur,
  ``Speaker recognition for multi-speaker conversations using x-vectors,'' in
  \emph{ICASSP 2019-2019 IEEE International Conference on Acoustics, Speech and
  Signal Processing (ICASSP)}.\hskip 1em plus 0.5em minus 0.4em\relax IEEE,
  2019, pp. 5796--5800.

\bibitem{garcia2020jhu}
D.~Garcia-Romero, A.~McCree, D.~Snyder, and G.~Sell, ``Jhu-hltcoe system for
  the voxsrc speaker recognition challenge,'' in \emph{ICASSP 2020-2020 IEEE
  International Conference on Acoustics, Speech and Signal Processing
  (ICASSP)}.\hskip 1em plus 0.5em minus 0.4em\relax IEEE, 2020, pp. 7559--7563.

\bibitem{desplanques20_interspeech}
B.~Desplanques, J.~Thienpondt, and K.~Demuynck, ``{ECAPA-TDNN: Emphasized
  Channel Attention, Propagation and Aggregation in TDNN Based Speaker
  Verification},'' in \emph{Proc. Interspeech 2020}, 2020, pp. 3830--3834.

\bibitem{thienpondt2021idlab}
J.~Thienpondt, B.~Desplanques, and K.~Demuynck, ``The idlab voxsrc-20
  submission: Large margin fine-tuning and quality-aware score calibration in
  dnn based speaker verification,'' in \emph{ICASSP 2021-2021 IEEE
  International Conference on Acoustics, Speech and Signal Processing
  (ICASSP)}.\hskip 1em plus 0.5em minus 0.4em\relax IEEE, 2021, pp. 5814--5818.

\bibitem{zhang21g_interspeech}
L.~Zhang, Q.~Wang, K.~A. Lee, L.~Xie, and H.~Li, ``{Multi-Level Transfer
  Learning from Near-Field to Far-Field Speaker Verification},'' in \emph{Proc.
  Interspeech 2021}, 2021, pp. 1094--1098.

\bibitem{heo2020clova}
H.~S. Heo, B.-J. Lee, J.~Huh, and J.~S. Chung, ``Clova baseline system for the
  voxceleb speaker recognition challenge 2020,'' \emph{2009.14153}, 2020.

\bibitem{zhang20ca_interspeech}
L.~Zhang, J.~Wu, and L.~Xie, ``{NPU Speaker Verification System for INTERSPEECH
  2020 Far-Field Speaker Verification Challenge},'' in \emph{Proc. Interspeech
  2020}, 2020, pp. 3471--3475.

\bibitem{2019Res2Net}
S.~Gao, M.~M. Cheng, K.~Zhao, X.~Y. Zhang, and P.~Torr, ``Res2net: A new
  multi-scale backbone architecture,'' \emph{IEEE Transactions on Pattern
  Analysis and Machine Intelligence}, vol.~PP, no.~99, pp. 1--1, 2019.

\bibitem{yu2020densely}
Y.-Q. Yu and W.-J. Li, ``Densely connected time delay neural network for
  speaker verification.'' in \emph{INTERSPEECH}, 2020, pp. 921--925.

\bibitem{thienpondt21_interspeech}
J.~Thienpondt, B.~Desplanques, and K.~Demuynck, ``{Integrating Frequency
  Translational Invariance in TDNNs and Frequency Positional Information in 2D
  ResNets to Enhance Speaker Verification},'' in \emph{Proc. Interspeech 2021},
  2021, pp. 2302--2306.

\bibitem{zhangdct}
L.~Zhang, Q.~Wang, and L.~Xie, ``Duality temporal-channel-frequency attention
  enhanced speaker,'' in \emph{IEEE 2011 workshop on automatic speech
  recognition and understanding}, no. CONF.\hskip 1em plus 0.5em minus
  0.4em\relax IEEE Signal Processing Society, 2021.

\bibitem{duta2020pyramidal}
I.~C. Duta, L.~Liu, F.~Zhu, and L.~Shao, ``Pyramidal convolution: Rethinking
  convolutional neural networks for visual recognition,'' \emph{2006.11538},
  2020.

\bibitem{vaswani2017attention}
A.~Vaswani, N.~Shazeer, N.~Parmar, J.~Uszkoreit, L.~Jones, A.~N. Gomez,
  {\L}.~Kaiser, and I.~Polosukhin, ``Attention is all you need,'' in
  \emph{Advances in neural information processing systems}, 2017, pp.
  5998--6008.

\bibitem{gulati20_interspeech}
A.~Gulati, J.~Qin, C.-C. Chiu, N.~Parmar, Y.~Zhang, J.~Yu, W.~Han, S.~Wang,
  Z.~Zhang, Y.~Wu, and R.~Pang, ``{Conformer: Convolution-augmented Transformer
  for Speech Recognition},'' in \emph{Proc. Interspeech 2020}, 2020, pp.
  5036--5040.

\bibitem{dai2021coatnet}
Z.~Dai, H.~Liu, Q.~V. Le, and M.~Tan, ``Coatnet: Marrying convolution and
  attention for all data sizes,'' \emph{2106.04803}, 2021.

\bibitem{yu2014mixed}
D.~Yu, H.~Wang, P.~Chen, and Z.~Wei, ``Mixed pooling for convolutional neural
  networks,'' in \emph{International conference on rough sets and knowledge
  technology}.\hskip 1em plus 0.5em minus 0.4em\relax Springer, 2014, pp.
  364--375.

\bibitem{cai18_odyssey}
W.~Cai, J.~Chen, and M.~Li, ``{Exploring the Encoding Layer and Loss Function
  in End-to-End Speaker and Language Recognition System},'' in \emph{Proc. The
  Speaker and Language Recognition Workshop (Odyssey 2018)}, 2018, pp. 74--81.

\bibitem{okabe18_interspeech}
K.~Okabe, T.~Koshinaka, and K.~Shinoda, ``{Attentive Statistics Pooling for
  Deep Speaker Embedding},'' in \emph{Proc. Interspeech 2018}, 2018, pp.
  2252--2256.

\bibitem{india19_interspeech}
M.~India, P.~Safari, and J.~Hernando, ``{Self Multi-Head Attention for Speaker
  Recognition},'' in \emph{Proc. Interspeech 2019}, 2019, pp. 4305--4309.

\bibitem{liu19f_interspeech}
Y.~Liu, L.~He, and J.~Liu, ``{Large Margin Softmax Loss for Speaker
  Verification},'' in \emph{Proc. Interspeech 2019}, 2019, pp. 2873--2877.

\bibitem{deng2020sub}
J.~Deng, J.~Guo, T.~Liu, M.~Gong, and S.~Zafeiriou, ``Sub-center arcface:
  Boosting face recognition by large-scale noisy web faces,'' in \emph{European
  Conference on Computer Vision}.\hskip 1em plus 0.5em minus 0.4em\relax
  Springer, 2020, pp. 741--757.

\bibitem{sun2020circle}
Y.~Sun, C.~Cheng, Y.~Zhang, C.~Zhang, L.~Zheng, Z.~Wang, and Y.~Wei, ``Circle
  loss: A unified perspective of pair similarity optimization,'' in
  \emph{Proceedings of the IEEE/CVF Conference on Computer Vision and Pattern
  Recognition}, 2020, pp. 6398--6407.

\bibitem{kingma2014adam}
D.~P. Kingma and J.~Ba, ``Adam: A method for stochastic optimization,''
  \emph{1412.6980}, 2014.

\bibitem{smith2017cyclical}
L.~N. Smith, ``Cyclical learning rates for training neural networks,'' in
  \emph{2017 IEEE winter conference on applications of computer vision
  (WACV)}.\hskip 1em plus 0.5em minus 0.4em\relax IEEE, 2017, pp. 464--472.

\bibitem{povey2011kaldi}
D.~Povey, A.~Ghoshal, G.~Boulianne, L.~Burget, O.~Glembek, N.~Goel,
  M.~Hannemann, P.~Motlicek, Y.~Qian, P.~Schwarz \emph{et~al.}, ``The kaldi
  speech recognition toolkit,'' in \emph{IEEE 2011 workshop on automatic speech
  recognition and understanding}, no. CONF.\hskip 1em plus 0.5em minus
  0.4em\relax IEEE Signal Processing Society, 2011.

\bibitem{brummer2013bosaris}
N.~Br{\"u}mmer and E.~De~Villiers, ``The bosaris toolkit: Theory, algorithms
  and code for surviving the new dcf,'' \emph{1304.2865}, 2013.

\end{thebibliography}
\end{document}